\documentclass{ifacconf}

\usepackage{graphicx}      
\usepackage{natbib}        
\usepackage{amsfonts}
\usepackage{amsmath}
\usepackage{color}
\usepackage{xparse}
\usepackage{algorithm}
\usepackage{multirow}
\usepackage{algpseudocode}
\usepackage{threeparttablex}
\makeatletter
\renewcommand{\ALG@name}{\scriptsize Alg.}
\makeatother
\algblockdefx[IF]{If}{EndIf}[1]{\footnotesize \algorithmicif\ #1\ \algorithmicthen}{\footnotesize\textbf{end}}
\algblockdefx[While]{While}{EndWhile}[1]{\footnotesize\algorithmicwhile\ #1}{\footnotesize\textbf{end}}
\algrenewcommand\algorithmicindent{0.5em}
\usepackage{setspace}
\let\Algorithm\algorithm
\renewcommand\algorithm[1][]{\Algorithm[#1]\setstretch{0.75}}
\NewDocumentCommand\Rm{mg}{%
    \IfNoValueTF{#2}
    {\mathbb{R}_{\max}^{#1}}
    {\mathbb{R}_{\max}^{#1 \times #2}}
}
\newcommand{\Rmax}{\mathbb{R}_{\max}}
\newcommand*{\QEDB}{\hfill\ensuremath{\square}}

\begin{document}
\begin{frontmatter}

\title{Symbolic Reachability Analysis of High Dimensional Max-Plus Linear Systems} 


\author[First]{Muhammad Syifa'ul Mufid} 
\author[Second]{Dieky Adzkiya} 
\author[First]{Alessandro Abate}

\address[First]{Department of Computer Science, University of Oxford, UK\\
 \emph{(}e-mail: \{muhammad.syifaul.mufid,alessandro.abate\}@cs.ox.ac.uk\emph{)}} 
 
\address[Second]{Department of Mathematics, ITS Surabaya, Indonesia
\\\emph{(}e-mail: dieky@matematika.its.ac.id\emph{)}}

\begin{abstract}                
This work discusses the reachability analysis (RA) of Max-Plus Linear (MPL) systems, a class of continuous-space, discrete-event models defined over the max-plus algebra. Given the initial and target sets, we develop algorithms to verify whether 
there exist trajectories of the MPL system that, starting from the initial set,
eventually reach the target set. 
We show that RA can be solved symbolically by encoding the MPL system, as well as initial and target sets into difference logic, 
and then checking the satisfaction of the resulting logical formula via an off-the-shelf satisfiability modulo theories (SMT) solver. 
The performance and scalability of the developed SMT-based algorithms are shown to clearly outperform state-of-the-art RA algorithms for MPL systems, newly allowing to investigate RA of high-dimensional MPL systems: 
the verification of models with more than 100 continuous variables shows the applicability of these techniques to MPL systems of industrial relevance. 
\end{abstract}

\begin{keyword}
max-plus linear systems, reachability analysis, piecewise-affine systems, difference-bound matrices, difference logic, satisfiability modulo theories
\end{keyword}

\end{frontmatter}

\section{Introduction}
Max-Plus Linear (MPL) systems are a subclass of discrete-event systems (DES) based on max-plus algebra which uses two binary operations, maximisation and addition. MPL systems are employed to describe synchronization without concurrency, and as such are widely used in transportation \citep{Heidergott} and manufacturing systems \citep{Aleksey}. A fundamental problem for DES is reachability analysis (RA): it investigates whether a certain set of states of the system is attainable from a given set of initial conditions. In the context of MPL systems, RA can be used to determine whether the trajectories of MPL system enter specific conditions that are deemed unsafe: for instance, in a railway network application \citep{Heidergott}, whether the delay between two consecutive train departures is ever greater than a given time inteval.  

The state-of-the-art approach for RA of MPL systems employs piecewise-affine (PWA) dynamics \citep{DiekyForward,DiekyBackward,Dieky2} and generates finite abstractions of MPL systems accordingly \citep{Dieky1}.  
Forward RA of MPL systems has been discussed in \citep{DiekyForward}.  
Given an initial set $X$, it computes the forward image of $X$ w.r.t. the underlying MPL system. 
Similarly, backward reachability of MPL systems is done by computing the inverse image of target set $Y$ \citep{DiekyBackward}, backwards in time. 
In \citep{DiekyForward,DiekyBackward}, both initial and target sets are assumed to be difference-bound matrices (DBMs) \citep{Dill} and the MPL dynamics are expressed as PWA models in the event domain. 

Whilst the approaches in \citep{DiekyForward,DiekyBackward,Dieky2} are scalable much beyond existing results based on simple algebraic operations, 
it is always desirable to push the envelope and to perform RA for MPL systems with ever larger number of variables (that is, with high continuous dimensions).  
In \citep{DiekyForward,DiekyBackward,Dieky2}, 
PWA systems are characterised by different spatial regions (PWA regions) and corresponding affine dynamics \citep{Sontag}. 
The translation of MPL systems into PWA dynamics, characterised by spatial regions and corresponding affine dynamics \citep{Sontag},  
has an exponential complexity \citep{Dieky1}: the number of PWA regions grows steeply as the dimension of MPL systems and the number of finite entries in the state matrix increase. Furthermore, the  forward and backward reach sets are characterised as unions of finitely many DBMs, the number of which grows exponentially w.r.t.\ the time horizon \citep{DiekyBackward,DiekyForward}. 

In order to attain scalability to really large MPL models, this paper proposes a symbolic approach to perform reachability analysis of MPL systems. Instead of computing reach sets explicitly, we use symbolic variables to encode the states of trajectories of MPL systems at each time horizon. Firstly, the MPL system as well as the initial and reach sets are translated into a formula that can be parsed by a satisfiability modulo theory (SMT) solver. An SMT problem deals with the satisfaction of a logical formula w.r.t. a given theory (e.g., linear arithmetics, or bit vectors) \citep{SMT}. Secondly, the satisfiability of the formula encoding a reachability problem is checked using an SMT solver. If the SMT solver reports ``satisfiable'' (resp. ``unsatisfiable''), then the target set is reachable (resp. not reachable) from an (resp. any) initial condition within the initial set. 

We have implemented the symbolic reachability analysis of MPL systems in C++, using the Z3 SMT solver \citep{Z3}. According to our numerical benchmark, the symbolic implementation is significantly faster than the state-of-the-art software tool. Furthermore, our implementation can solve the reachability analysis of 100-dimensional MPL systems within reasonable time and memory requirements: these results render RA of MPL systems newly applicable to industrial-sized models. 

The paper is structured as follows. Section 2 introduces the basic notions of MPL systems and the brief summary of reachability analysis based on reach sets computation. 
Section 3 consists of the brief definition of SMT and the main contribution of this paper. The computational benchmarks are provided in Section 4. Finally, we conclude the paper with Section 5.

\section{Model and Preliminaries}
\subsection{Max-Plus Linear Systems}
\indent Max-plus algebra is an algebraic structure $(\Rmax,\oplus,\otimes)$ where
$\Rmax:=\mathbb{R}\cup\{\varepsilon:=-\infty\}$ and 
\[
a\oplus b:=\max\{a,b\}~\text{and}~a\otimes b:=a+b
\]
for all $a,b\in \Rmax$. 
These operations can be extended to matrices and vectors, as follows:
\begin{align}
\nonumber [\alpha\otimes A](i,j)&=\alpha + A(i,j),\\
\nonumber [A\oplus B](i,j)&=A(i,j) \oplus B(i,j),\\
\nonumber [A\otimes C](i,j)&=\bigoplus_{k=1}^n A(i,k)\otimes C(k,j),
\end{align}
where $A,B\in \Rm{m}{n}, C\in \Rm{n}{p}$ and $\alpha\in \Rmax$. Given $A\in\Rm{n}{n}$ and $r\in \mathbb{N}$, $A^{\otimes r}$ denotes $A\otimes\ldots\otimes A$ ($r$ times). 

A Max-Plus Linear (MPL) system is defined as 
\begin{equation}
\textbf{x}(k)=A\otimes \textbf{x}(k-1), ~~k=1,2,\ldots
\label{mpl}
\end{equation}
where $A\in \Rm{n}{n}$ is the system matrix and vector $\textbf{x}(k)=[x_1(k)~\ldots~ x_n(k)]^\top$ is the state variables \citep{Baccelli}. In applications, $\textbf{x}$ represents the time stamps of the discrete events, while $k$ corresponds to the event counter. Hence, it is more convenient to take $\mathbb{R}^n$ as the state space and $A$ to be a regular matrix, i.e. there exists at least one finite element in each row of $A$ \citep{Heidergott}.

\begin{defn}\citep{Baccelli}. The precedence graph of $A\in\Rm{n}{n}$, denoted by $\mathcal{G}(A)$, is a weighted directed graph with nodes $1,\ldots,n$ and an edge from $j$ to $i$ with weight $A(i,j)$ for each $A(i,j)\neq \varepsilon$.\QEDB
\end{defn}
The readers are referred to \citep{Baccelli} for more detailed descriptions about $\mathcal{G}(A)$ including the notions of \textit{strongly connected} and \textit{critical circuit}.
\begin{defn}\citep{Baccelli}. A matrix $A\in \Rm{n}{n}$ is called \textit{irreducible} if $\mathcal{G}(A)$ is strongly connected.\QEDB
\end{defn}


Each irreducible matrix $A\in\Rmax^{n\times n}$ admits a unique max-plus eigenvalue $\lambda\in\mathbb{R}$, which corresponds to the average weight of critical circuit in $\mathcal{G}(A)$. Furthermore, $A$ satisfies the so-called transient condition:  
\begin{prop}\citep{Baccelli}. For an irreducible matrix $A\in\Rmax^{n\times n}$ and its max-plus eigenvalue $\lambda\in\mathbb{R}$, there exist $k_0,c\in\mathbb{N}$ such that $A^{\otimes(k+c)}=\lambda c\otimes A^{\otimes k}$ for all $k\geq k_0$. The smallest such $k_0$ and $c$ are called the \textit{transient} and the \textit{cyclicity} of $A$, respectively. \QEDB
\label{trans}
\end{prop}
\subsection{Difference-Bound Matrices}
Difference-Bound Matrices (DBMs) are defined as the intersection of sets defined by the difference of two variables.
\begin{defn}\citep{Dill}. A DBM in $\mathbb{R}^n$ is the intersection of sets defined by $x_i-x_j\sim_{i,j} d_{i,j}$, where $\sim_{i,j}~\in \{>,\geq\}$ and $d_{i,j}\in \Rmax$ for $0\leq i,j\leq n$. The value of special variable $x_0$ is always equal to 0.\QEDB
\label{DBM_def}
\end{defn}
The variable $x_0$ is used to represent inequalities with a single variable: $x_i\geq \alpha$ can be written as $x_i-x_0\geq \alpha$. Unless otherwise stated, in this work we assume that DBM does not contain any inequality with a single variable. 
Some operations can be applied to DBMs, such as intersection, canonical-form representation, emptiness checking, image and inverse image w.r.t. an affine dynamic \citep{Dieky1,Mufid2018}. 
\subsection{Piecewise-Affine Systems}
Piecewise-Affine (PWA) systems \citep{Sontag} are defined by partitioning the input-state space into several domains characterized by polyhedra. Each domain, or PWA region, is associated with an affine function. It is shown in \citep{Heemels} that every MPL system can be transformed into a PWA system.  For $A$ in \eqref{mpl}, the PWA regions are generated from $\textbf{g}=(g_1,\ldots,g_n)\in \{1,\ldots,n\}^n$, 
which 
satisfies
$A(i,g_i)\neq \varepsilon $ for $ 1\leq i\leq n$. The region corresponding to $\textbf{g}$ is
\begin{equation}
\mathsf{R}_\textbf{g} \!= \!\bigcap_{i=1}^n\bigcap_{j=1}^n\left\{\textbf{\text{x}}\in \mathbb{R}^n|x_{g_i}\!- x_j\geq  A(i,j)\!-\!A(i,g_i)\right\}. 
\label{pwa}
\end{equation}
Notice that, $\mathsf{R}_\textbf{g}$ is a DBM. The emptiness checking of $\mathsf{R}_\textbf{g}$ can be done using Floyd-Warshall algorithm which has cubic complexity w.r.t. its dimension \citep{Floyd}. The affine dynamics for a non-empty $\mathsf{R}_\textbf{g}$ is
\begin{equation}
x_i(k)=x_{g_i}(k-1)+A(i,g_i),~~ i=1,\ldots,n.
\label{af}
\end{equation}
\begin{exmp}
Consider a $2\times 2$ MPL system \eqref{mpl} where 
\begin{equation}
A=\begin{bmatrix}
2 & ~5\\ 3 & ~3
\end{bmatrix}.
\label{mpl_ex}
\end{equation}
The resulting PWA regions are $\mathsf{R}_{(1,1)}=\{\textbf{x}\in \mathbb{R}^2\mid x_1-x_2\geq 3\}$, $\mathsf{R}_{(2,1)}=\{\textbf{x}\in \mathbb{R}^2\mid0\leq x_1-x_2\leq  3\}$, and $\mathsf{R}_{(2,2)}=\{\textbf{x}\in \mathbb{R}^2\mid x_1-x_2\leq 0\}$. The corresponding affine dynamics is
\[
\begin{bmatrix}
x_1(k)\\
x_2(k)
\end{bmatrix}
 = 
  \begin{cases}
\begin{bmatrix}
x_1(k-1)+2\\x_1(k-1)+3
\end{bmatrix},&\text{if}~\textbf{x}(k-1)\in \mathsf{R}_{(1,1)}, \\
\begin{bmatrix}
x_2(k-1)+5\\x_1(k-1)+3
\end{bmatrix},&\text{if}~\textbf{x}(k-1)\in \mathsf{R}_{(2,1)}, \\
\begin{bmatrix}
x_2(k-1)+3\\x_2(k-1)+3
\end{bmatrix},&\text{if}~\textbf{x}(k-1)\in \mathsf{R}_{(2,2)}.
  \end{cases}
\]
\QEDB
\end{exmp}

\section{Explicit Reachability Analysis of Max-Plus Linear Systems}
Suppose we have an MPL system \eqref{mpl} and ${X},Y\subseteq\mathbb{R}^n$ as the initial and target sets, respectively.
The set $Y$ is \textit{reachable at time} $k$ from $X$ if there exist $\textbf{x}(0) \in X$ such that $\textbf{x}(k)\in Y$, where $\textbf{x}(k)$ is computed recursively by \eqref{mpl} from $\textbf{x}(0)$. 
The existing approach for solving reachability analysis (RA) of MPL systems is by computing forward and backward reach sets of MPL systems \citep{DiekyBackward,DiekyForward,Dieky2}. 

Given an initial set $X$, the forward reach set $X_k$ is recursively defined as 
\begin{equation}
X_k=\mathtt{Im}(X_{k-1})=\{A\otimes \textbf{x}\mid \textbf{x} \in X_{k-1}\},
\label{fw}
\end{equation}
where $X_0=X$. Likewise, from the target set $Y$, the backward reach set $Y_{-k}$ is defined as 
\begin{equation}
  Y_{k-1}=\mathtt{Im}^{-1}(Y_{k})=\{\textbf{y}\in\mathbb{R}^n\mid A\otimes \textbf{y} \in Y_{k}\},   
\label{bw}
\end{equation}
where $Y_0=Y$. The initial and target states are assumed to be non-empty DBMs. The forward and backward reach sets can be computed using one-shot procedures as follows:
\begin{equation}
X_k=\{A^{\otimes k}\otimes \textbf{x}\mid x \in X_{0}\},
\label{fw2}
\end{equation}
and 
\begin{equation}
Y_{-k}=\{\textbf{y}\in\mathbb{R}^n\mid A^{\otimes k}\otimes \textbf{y} \in Y_{0}\}.
\label{bw2}
\end{equation}

To compute forward and backward reach sets, one needs to represent an MPL system \eqref{mpl} as a PWA model. The steps to compute $X_k$ are explained in \citep{DiekyForward} and involve image computation of DBMs w.r.t.\ the affine dynamics. On the other hand, the inverse image computation of DBMs w.r.t. affine dynamics is used to compute $Y_{-k}$ \citep{DiekyBackward}. It has been shown in \citep{DiekyBackward,DiekyForward} that both forward and backward reach sets are a union of finitely many DBMs. Notice that, $X_k\neq \emptyset$ for $k\geq 0$. However, it is possible that there is {an} $l>0$ such that {$Y_{-k}=\emptyset$} for all $k\geq l$.

Algorithms 1-4 illustrate ways to perform RA of MPL systems by means of the computation of forward and backward reach sets up to a given bound $N\in \mathbb{N}$. In Algorithms 2 and 4, one needs to generate the PWA system for $A^{\otimes k}$ for each iteration $k$ - notice that this ``one-shot'' implementation does not simply compute the reach set at the final time $N$, as it still runs over the entire time horizon; later we will reason about the benefit of such implementation versus the ``sequential'' Algorithms 1 and 3. For Algorithms 3-4, if the backward reach set $Y_{-k}=\emptyset$ then the algorithms are terminated at the $k^\text{th}$ iteration with $\texttt{false}$ as the output. 

If the output of Algorithm 1-4 is $\texttt{true}$, 
then $Y$ is reachable from $X$, 
otherwise $Y$ is not reachable from $X$, within the given time bound $N$. 
Given a negative outcome from above, 
in general we cannot conclude that $Y$ is not reachable from $X$ within time bounds greater than $N$. 
However, for irreducible MPL systems we can prove that there exists a \emph{completeness threshold} \citep{CT} $N^{\ast}\in \mathbb{N}$ for Algorithms 1-4. 
This notion is widely used in the model checking literature and applies to RA. 
Such a scalar is the maximum iteration that is sufficient for the termination of an algorithm: 
e.g. for Algorithms 1-4, if $Y$ is not reachable from $X$ up to bound $N^\ast$, then $Y$ is surely also not reachable from $X$ within any larger bound $N>N^{\ast}$.  
So quite importantly, finding a completeness threshold ensures the completeness of RA procedures. 

We show that the completeness threshold is related to transient and cyclicity of irreducible MPL systems. It is important to note that the transient of an irreducible MPL system is not linear w.r.t. its dimension (a small dimensional MPL system may have a relatively large transient).
\begin{prop}
\label{max_iter}
If $A\in \Rm{n}{n}$ is irreducible then the completeness threshold for Algorithms 1-4 is $k_0+c-1$, where $k_0$ and $c$ are the transient and cyclicity of $A$, respectively. 
\end{prop}
\begin{pf}
It suffices to prove the completeness threshold for Algorithms 1 and 3. Suppose we have forward reach sets $X_0,X_1,\ldots,$ where $X_k=\mathtt{Im}(X_{k-1})$. By Proposition \ref{trans}, for $k\geq k_0$ we have $A^{\otimes(k+c)}=\lambda c\otimes A^{\otimes k}$, which implies $\textbf{x}\in X_{k}$ iff $\lambda c\otimes \textbf{x}\in X_{k+c}$. Recall that the forward reach sets are in general unions of DBMs. Furthermore, DBMs are not affected by shifting operations\footnote{Given a DBM $D$ and $\alpha\in \mathbb{R}$, $\alpha\otimes D =\{\alpha\otimes \textbf{x}\mid \textbf{x}\in D\}= D$.}. Consequently, $X_{k+c}=X_k$ for $k\geq k_0$. From here, we can conclude that we only need to consider a bound before reaching the periodicity, i.e. $k_0+c-1$. Now, suppose we have non-empty backward reach sets $Y_0,Y_{-1},\ldots,$ where $Y_{k-1}=\mathtt{Im}^{-1}(Y_{k})$. Similarly, by Proposition \ref{trans}, we have $Y_{-(k+c)}=Y_{-k}$ for $k\geq k_0$ which leads to the same conclusion as previous one.\QEDB
\end{pf}

By Proposition \ref{max_iter}, we can conclude that RA of irreducible MPL system is decidable, provided that the initial and target sets are DBMs.
\vspace*{-1ex}\\
\begin{minipage}{0.23\textwidth}
\begin{algorithm}[H]
	\footnotesize
    \centering
    \hspace*{-12ex}\textbf{Inputs:} $A\in \Rm{n}{n}$,\\
    \hspace*{-0.5ex}  initial set $X$,\\
    \hspace*{-1ex} target set $Y$,\\
    \hspace*{-8ex} $N\in \mathbb{N}$\\
    \hspace*{-13ex}\textbf{Output:} boolean
    \caption{\scriptsize RA (forward)}
    \label{alg1}
    \begin{algorithmic}[1]
        \State $reach\gets \texttt{false}$
        \State $X_0\gets X$
        \State generate PWA system of $A$
        \State $k\gets 1$
        \While{$k\leq N$}
        \State compute $X_k$ by \eqref{fw}
        \If{$X_k\cap Y\neq \emptyset$}
        \State $reach\gets \texttt{true}$
        \State \textbf{break}
        \EndIf
        \State $k\gets k+1$
        \EndWhile
        \State \textbf{return} $reach$
    \end{algorithmic}
\end{algorithm}
\end{minipage}
\begin{minipage}{0.25\textwidth}
\begin{algorithm}[H]
	\footnotesize
    \centering
    \hspace*{-12ex}\textbf{Inputs:} $A\in \Rm{n}{n}$,\\
    \hspace*{-0.5ex}  initial set $X$,\\
    \hspace*{-1ex} target set $Y$,\\
    \hspace*{-8ex} $N\in \mathbb{N}$\\
    \hspace*{-13ex}\textbf{Output:} boolean
    \caption{\scriptsize RA (one-shot forward)}
    \label{alg2}
    \begin{algorithmic}[1]
        \State $reach\gets \texttt{false}$
        \State $X_0\gets X$
        \State $k\gets 1$
        \While{$k\leq N$}
        \State generate PWA system of $A^{\otimes k}$
        \State compute $X_k$ by \eqref{fw2}
        \If{$X_k\cap Y\neq \emptyset$}
        \State $reach\gets \texttt{true}$
        \State \textbf{break}
        \EndIf
        \State $k\gets k+1$
        \EndWhile
        \State \textbf{return} $reach$
    \end{algorithmic}
\end{algorithm}
\end{minipage}
\vspace*{-1ex}\\
\begin{minipage}{0.23\textwidth}
\begin{algorithm}[H]
	\footnotesize
    \centering
    \hspace*{-12ex}\textbf{Inputs:} $A\in \Rm{n}{n}$,\\
    \hspace*{-0.5ex}  initial set $X$,\\
    \hspace*{-1ex} target set $Y$,\\
    \hspace*{-8ex} $N\in \mathbb{N}$\\
    \hspace*{-13ex}\textbf{Output:} boolean
    \caption{\scriptsize RA (backward)}
    \label{alg3}
    \begin{algorithmic}[1]
        \State $reach\gets \texttt{false}$
        \State $Y_0\gets Y$
        \State generate PWA system of $A$
        \State $k\gets 1$
        \While{$k\leq N$}
        \State compute $Y_{-k}$ by \eqref{bw}
        \If{$Y_{-k}= \emptyset$}
        \State \textbf{break}
        \EndIf
        \If{$Y_{-k}\cap X\neq \emptyset$}
        \State $reach\gets \texttt{true}$
        \State \textbf{break}
        \EndIf
        \State $k\gets k+1$
        \EndWhile
        \State \textbf{return} $reach$
    \end{algorithmic}
\end{algorithm}
\end{minipage}
\begin{minipage}{0.25\textwidth}
\begin{algorithm}[H]
	\footnotesize
    \centering
    \hspace*{-12ex}\textbf{Inputs:} $A\in \Rm{n}{n}$,\\
    \hspace*{-0.5ex}  initial set $X$,\\
    \hspace*{-1ex} target set $Y$,\\
    \hspace*{-8ex} $N\in \mathbb{N}$\\
    \hspace*{-13ex}\textbf{Output:} boolean
    \caption{\scriptsize RA (one-shot backward)}
    \label{alg4}
    \begin{algorithmic}[1]
        \State $reach\gets \texttt{false}$
        \State $Y_0\gets Y$
        \State $k\gets 1$
        \While{$k\leq N$}
        \State generate PWA system of $A^{\otimes k}$
        \State compute $Y_{-k}$ by \eqref{bw2}
        \If{$Y_{-k}= \emptyset$}
        \State \textbf{break}
        \EndIf
        \If{$Y_{-k}\cap X\neq \emptyset$}
        \State $reach\gets \texttt{true}$
        \State \textbf{break}
        \EndIf
        \State $k\gets k+1$
        \EndWhile
        \State \textbf{return} $reach$
    \end{algorithmic}
\end{algorithm}
\end{minipage}

We provide an example of reachability analysis of MPL systems via reach sets computation.
\begin{exmp}
\label{exmp_RA}
With the preceding MPL system in Example 5, we define the initial and target sets respectively as $X=\{\textbf{x}\in \mathbb{R}^2\mid x_1-x_2\geq 3\}$ and $Y=\{\textbf{x}\in \mathbb{R}^2\mid x_1-x_2\geq 5\}$. One could check that the transient and cyclicity of \eqref{mpl_ex} are $k_0=c=2$ and therefore the completeness threshold is $N^\ast=3$. 

Leaving details aside, the forward reach sets are $X_1=\{\textbf{x}\in \mathbb{R}^2\mid x_1-x_2= -1\},X_2=\{\textbf{x}\in \mathbb{R}^2\mid x_1-x_2= 0\}$, and $X_3=\{\textbf{x}\in \mathbb{R}^2\mid x_1-x_2= 2\}$. As $X_i\cap Y=\emptyset$ for $i=1,2,3$, we can conclude that $Y$ is not reachable from $X$. By backward reach set computation, we have $Y_1=\emptyset$ which leads to the same conclusion. \QEDB

\end{exmp}

There are a few elements contributing to the computational bottleneck (time and memory requirements) of this approach.
First of all, the number of regions in the PWA systems depends on the size of state matrix and on the number of finite entries in the matrix. 
The worst-case complexity of generating the PWA system via \eqref{pwa} is $\mathcal{O}(n^{n+3})$ \citep{Dieky1}.  Furthermore, the reachable set and backward reachable set are a union of finitely many DBMs. In the worst case, the number of DBMs grows exponentially with the time horizon.

As shown in \citep{DiekyForward}, the worst-case complexity to generate the sequential (resp. one-shot) reach sets up to bound $N$ is $\mathcal{O}(\sum_{k=0}^{N-1}|X_k|\cdot n^{n+3})$ (resp. $\mathcal{O}((\lfloor\log_2{N}\rfloor+|X_0|)\cdot n^{n+3})$) , where $|X_k|$ represents the number of DBMs in $X_k$. Similarly, the complexity for backward reach sets computations are $\mathcal{O}(\sum_{k=0}^{N-1}|Y_k|\cdot n^{n+3})$ (for sequential) and $\mathcal{O}((\lfloor\log_2{N}\rfloor+|Y_0|)\cdot n^{n+3})$ (for one-shot). Surely, the one-shot procedures are more efficient than the sequential ones.





\section{Symbolic Reachability Analysis of Max-Plus Linear Systems}
\subsection{Satisfiability Modulo Theories}
Satisfiability Modulo Theories (SMT) deal with the problem of determining the satisfaction of a first-order logical formula w.r.t. some logical theory background, 
such as Boolean logic (which generalises SAT theory), bit-vectors, real and integer arithmetics, and so on \citep{SMT}. For instance, the following formula 
\begin{equation}
\label{smt}
\nonumber
(x\geq 0) \wedge (y<2) \wedge (x-y< -1)
\end{equation}
has solutions for $x,y\in \mathbb{R}$ but no solution for $x,y\in \mathbb{Z}$. 
In general, an SMT formula may contain conjunctions ($\wedge$), disjunctions ($\vee$), and quantifiers $(\exists,\forall)$. An SMT solver reports whether the given formula is satisfiable or not satisfiable. For the former case, it usually also provides a \textit{model}, i.e. a satisfying assignment for the formula. 

SMT has grown into a very active research subject: it has a standardised library and a collection of benchmarks developed by the SMT community \citep{SMT-LIB}, as well as an annual international competition for SMT solvers \citep{SMT-COMP}. 
As a result, there are several powerful SMT solvers, 
such as MATHSAT5 \citep{MATHSAT}, Yices 2.2 \citep{Yices}, and Z3 \citep{Z3}. 
Applications of SMT-solving arise on supervisory control of discrete-event systems \citep{Shoaei}, verification of neural networks \citep{Reluplex}, optimization \citep{Yi}, and beyond. 

\subsection{SMT-Based Reachability Analysis of MPL systems}
This section discusses new procedures to solve RA of MPL systems using SMT-solving. 
We use quantifier-free \textit{difference logic} as the underlying logical theory for SMT. 

\begin{defn}[Difference logic, \citep{Cotton}]~\\Let $\mathcal{B}=\{\texttt{b}_1,\ldots,\texttt{b}_m\}$ and $\mathcal{V}=\{\texttt{x}_1,\ldots,\texttt{x}_n\}$ be sets of Boolean and real-valued variables, respectively. The set of atomic formulae of $DL(\mathcal{B},\mathcal{V})$ consists of Boolean variables in $\mathcal{B}$ and of inequalities with the form $\texttt{x}_i-\texttt{x}_j\sim c$, where $\sim\hspace*{0.5ex} \in \{>,\geq\}$ and $c\in \mathbb{R}$.\QEDB
\label{diff_logic}
\end{defn}
For instance, both $f_1=(\texttt{x}_1-\texttt{x}_2\geq 1)\rightarrow (\texttt{x}_1-\texttt{x}_3>1)$ and $f_2=((\texttt{x}_1-\texttt{x}_2>2)\wedge \texttt{b}_1)\leftrightarrow(\neg(\texttt{x}_2-\texttt{x}_3\geq 0)\vee \texttt{b}_2)$ are formulae in difference logic. In this work, we only consider formulae in difference logic where the Boolean variables do not appear, as in $f_1$. 
Interestingly, notice that any DBM is a formula in difference logic, where Boolean connectives are exclusively conjunctions $(\wedge)$: 
as such, the non-emptiness of a DBM is equivalent to the satisfiability of its corresponding difference logic formula. 

We show that operations in max-plus algebra can be expressed as formulae in difference logic. 
\begin{prop}
Given real-valued variables $\texttt{x}_1,\ldots,\texttt{x}_n$ and real scalars $a_1,\ldots,a_n$, the equation
$\texttt{x}^\prime=\bigoplus_{i=1}^n (\texttt{x}_i\otimes a_i) $ is equivalent to 
\begin{equation}
\left(\bigwedge_{i=1}^n(\texttt{x}^\prime-\texttt{x}_i\geq a_i)\right )\wedge \left(\bigvee_{i=1}^n(\texttt{x}^\prime-\texttt{x}_i= a_i)\right ). 
\label{mpl_rdl}
\end{equation}
\end{prop}
\begin{pf} 
The difference logic formula \eqref{mpl_rdl} asserts that: 1) $\forall i~\texttt{x}^\prime \geq \texttt{x}_i+a_i$ and 2) $\exists i~\texttt{x}^\prime =\texttt{x}_i+a_i$. 
From both conditions, it is straigthforward that $\texttt{x}^\prime$ can be expressed as $ \max\{\texttt{x}_1+a_1,\ldots,\texttt{x}_n+a_n\}$.\QEDB
\end{pf}
For the rest of the paper, $\mathcal{V}^{(k)}=\{\texttt{x}^{(k)}_1,\ldots,\texttt{x}^{(k)}_n\}$ denotes the set of variables encompassing the states of the MPL system in \eqref{mpl} at the $k^\text{th}$ event. 
Via Proposition \ref{mpl_rdl}, the MPL system in \eqref{mpl} can be expressed as a formula in difference logic as follows: 
\begin{equation}
\mathtt{Im}(\mathcal{V}^{(k-1)},\mathcal{V}^{(k)})=\bigwedge_{i=1}^n(\texttt{ge}_i\wedge \texttt{eq}_i), 
\label{mpl_smt}
\end{equation}
where $$\displaystyle\texttt{ge}_i=\bigwedge_{j\in \mathtt{fin}_i} (\texttt{x}^{(k)}_i-\texttt{x}^{(k-1)}_j\geq A(i,j)),$$
$$\texttt{eq}_i=\bigvee_{j\in \mathtt{fin}_i} (\texttt{x}^{(k)}_i-\texttt{x}^{(k-1)}_j=A(i,j)),$$
and $\mathtt{fin}_i$ is a set containing the indices of the finite elements of $A(i,\cdot)$. 

Consequently, the following SMT formula 
\begin{equation}
T=\bigwedge_{k=1}^N \mathtt{Im}(\mathcal{V}^{(k-1)},\mathcal{V}^{(k)})
\label{T1}
\end{equation}
corresponds to a \textit{symbolic representation} of the states of the trajectory of the MPL system in \eqref{mpl} for $k=1,\ldots,N$. 

It follows that the reachability of the target set $Y$ from the initial set $X$ up to bound $N$ can be equivalently expressed as the satisfiability of the SMT formula 
\begin{equation}
X^{(0)}\wedge T \wedge Y^{(N)}, 
\label{test1}
\end{equation}
where $X^{(0)}$ (resp. $Y^{(N)}$) is the difference logic representation for $X$ (resp. $Y$) over $\mathcal{V}^{(0)}$ (resp. $\mathcal{V}^{(N)}$). 

Furthermore, the one-shot approach to reachability analysis can be formulated symbolically from \eqref{T1} as follows:  
instead of using $N$ conjuncts, \eqref{T1} can be expressed by 
\[T = \mathtt{Im}^N(\mathcal{V}^{(0)},\mathcal{V}^{(1)}), \]  
where $\mathtt{Im}^N$ is generated from \eqref{mpl_smt} for matrix $A^{\otimes N}$. 
Accordingly, the formula \eqref{test1} is changed into 
\begin{equation}
X^{(0)}\wedge  T \wedge  Y^{(1)}.
\label{test2}
\end{equation}
{\begin{exmp} 
With the previous example of explicit reachability analysis in Example \ref{exmp_RA}, the corresponding formula \eqref{test1} for $N=3$ is
\[
(\texttt{x}^{(0)}_1-\texttt{x}^{(0)}_2\geq 3)\wedge T_1\wedge T_2 \wedge T_3 \wedge (\texttt{x}^{(3)}_1-\texttt{x}^{(3)}_2\geq 5),
\]
where 
\begin{align*}
T_k=& (\texttt{x}^{(k)}_1-\texttt{x}^{(k-1)}_1\!\geq\! 2)\!\wedge\! (\texttt{x}^{(k)}_1-\texttt{x}^{(k-1)}_2\!\geq\! 5)\wedge \\
&(\!(\!\texttt{x}^{(k)}_1-\texttt{x}^{(k-1)}_1\!=\! 2)\!\vee\! (\texttt{x}^{(k)}_1-\texttt{x}^{(k-1)}_2\!=\! 5\!)\!)\wedge\\
&(\texttt{x}^{(k)}_2-\texttt{x}^{(k-1)}_1\!\geq\! 3)\!\wedge \!(\texttt{x}^{(k)}_2-\texttt{x}^{(k-1)}_2\!\geq\! 3)\wedge \\
&(\!(\!\texttt{x}^{(k)}_2-\texttt{x}^{(k-1)}_1\!=\! 3)\!\vee \!(\texttt{x}^{(k)}_2-\texttt{x}^{(k-1)}_2\!=\! 3)\!).
\end{align*}
On the other hand, the one-shot version formula \eqref{test2} is
\begin{align*}
& (\texttt{x}^{(0)}_1-\texttt{x}^{(0)}_2\!\geq\! 3)\!\wedge \!(\texttt{x}^{(1)}_1-\texttt{x}^{(0)}_1\!\geq\! 11)\!\wedge\! (\texttt{x}^{(1)}_1-\texttt{x}^{(0)}_2\!\geq\! 13)\wedge \\
&(\!(\texttt{x}^{(1)}_1-\texttt{x}^{(0)}_1\!=\! 11)\!\vee\! (\texttt{x}^{(1)}_1-\texttt{x}^{(0)}_2\!=\! 13)\!)\!\wedge\! (\texttt{x}^{(1)}_2-\texttt{x}^{(0)}_1\!\geq\! 11)\wedge\\
& (\texttt{x}^{(1)}_2-\texttt{x}^{(0)}_2\!\geq\! 11)\!\wedge\! (\!(\!\texttt{x}^{(1)}_2-\texttt{x}^{(0)}_1\!=\! 11)\!\vee \!(\texttt{x}^{(1)}_2-\texttt{x}^{(0)}_2\!=\! 11)\!)\wedge\\
&(\texttt{x}^{(1)}_1-\texttt{x}^{(1)}_2\!\geq\! 5),
\end{align*}
where the first (resp. last) conjunct corresponds to the initial (resp. target) set in Example \ref{exmp_RA}. \QEDB
\end{exmp}

Algorithms 5-6 illustrate the SMT-based adaptation of Algorithms 1-2. The function $\mathsf{fresh}\_\mathsf{var}(k,n)$ generates a set of $n$ real-valued variables for bound $k$. 
$F$ is a \textit{program vector} (not be confused with vectors in linear or max-plus algebra) containing SMT formulae as in \eqref{test1}. The command $\mathsf{push}\_\mathsf{back}$ adds a formula into $F$ from the back while $\mathsf{pop}\_\mathsf{back}$ removes the last one. For $i\geq 0$, $F[i]$ is the $(i+1)^\text{th}$ element of $F$ (from the back).

At the start, both $X$ and $Y$ are expressed as difference logic over $\mathcal{V}^{(0)}$. 
The function $Y.\mathsf{subs}(\mathcal{V}^{(k-1)},\mathcal{V}^{(k)})$ substitutes each appearance of $\texttt{x}_i^{(k-1)}$ in $Y$ with $\texttt{x}_i^{(k)}$. 
The non-emptiness checking of a union of DBMs in line 7 of Algorithms 1-2 is now formulated as the satisfiability checking of a difference logic formula (line 12 and 13 of Algorithm 5 and 6, respectively), where $\wedge F$ stands for $\wedge_{0\leq i<|F|} F[i]$. 
The $\mathsf{check}$ function is implemented by an SMT solver, where $\mathsf{check}(\wedge F)=\texttt{true}$ means that $\wedge F$ is satisfiable. 

In lines 8-11 of Algorithm \ref{alg5}, a difference logic formula for \eqref{mpl_smt} and the target set over $\mathcal{V}^{(k)}$ are added to $F$ for each iteration $k$. 
If the condition in line 12 is not fulfilled then the last element of $F$ (i.e., $Y$) is removed. 
For Algorithm \ref{alg6}, the number of elements in $F$ is three for each iteration. 
In line 6, we set a temporary element for $F[1]$, which will be changed at each iteration (line 12).

\begin{minipage}{0.24\textwidth}
\begin{algorithm}[H]
	\footnotesize
    \centering
    \hspace*{-12ex}\textbf{Inputs:} $A\in \Rm{n}{n}$,\\
    \hspace*{-0.5ex}  initial set $X$,\\
    \hspace*{-1ex} target set $Y$,\\
    \hspace*{-8ex} $N\in \mathbb{N}$,\\
    \hspace*{-13ex}\textbf{Output:} boolean
    \caption{\scriptsize SMT-RA (forward)}
    \label{alg5}
    \begin{algorithmic}[1]
        \State $reach\gets \texttt{false}$
        \State $\mathcal{V}^{(0)}\gets \mathsf{fresh}\_\mathsf{var}(0,n)$
        \State $F\gets \emptyset$ \Comment{{\scriptsize empty vector}}
        \State $F.\mathsf{push}\_\mathsf{back}(X)$       
        \State $k\gets 1$
        \While{$k\leq N$}
        \State $\mathcal{V}^{(k)}\gets \mathsf{fresh}\_\mathsf{var}(k,n)$
        \State $mpl\gets \mathtt{Im}(\mathcal{V}^{(k-1)},\mathcal{V}^{(k)})$
        \State $F.\mathsf{push}\_\mathsf{back}(mpl)$
        \State $Y.\mathsf{subs}(\mathcal{V}^{(k-1)},\mathcal{V}^{(k)})$
        \State $F.\mathsf{push}\_\mathsf{back}(Y)$
        \If{$\mathsf{check}(\wedge F)\!=\!\texttt{true}$}
        \State $reach\gets \texttt{true}$
        \State \textbf{break}
        \EndIf
        \State $F.\mathsf{pop}\_\mathsf{back}()$
        \State $k\gets k+1$
        \EndWhile
        \State \textbf{return} $reach$
    \end{algorithmic}
\end{algorithm}
\end{minipage}
\begin{minipage}{0.24\textwidth}
\begin{algorithm}[H]
	\footnotesize
    \centering
    \hspace*{-12ex}\textbf{Inputs:} $A\in \Rm{n}{n}$,\\
    \hspace*{-0.5ex}  initial set $X$,\\
    \hspace*{-1ex} target set $Y$,\\
    \hspace*{-8ex} $N\in \mathbb{N}$,\\
    \hspace*{-13ex}\textbf{Output:} boolean
    \caption{\scriptsize SMT-RA (one-shot forward)}\label{alg6}
    \begin{algorithmic}[1]
        \State $reach\gets \texttt{false}$
        \State $\mathcal{V}^{(0)}\gets \mathsf{fresh}\_\mathsf{var}(0,n)$
        \State $\mathcal{V}^{(1)}\gets \mathsf{fresh}\_\mathsf{var}(1,n)$
        \State $F\gets \emptyset$ \Comment{{\scriptsize empty vector}}
        \State $F.\mathsf{push}\_\mathsf{back}(X)$  
        \State $F.\mathsf{push}\_\mathsf{back}(\texttt{true})$ 
        \State $Y.\mathsf{subs}(\mathcal{V}^{(0)},\mathcal{V}^{(1)})$ 
        \State $F.\mathsf{push}\_\mathsf{back}(Y)$
        \State $k\gets 1$
        \While{$k\leq N$}
        \State $mpl\gets \mathtt{Im}^k(\mathcal{V}^{(0)},\mathcal{V}^{(1)})$
        \State $F[1]\gets mpl$
        \If{$\mathsf{check}(\wedge F)=\texttt{true}$}
        \State $reach\gets \texttt{true}$
        \State \textbf{break}
        \EndIf
        \State $k\gets k+1$
        \EndWhile
        \State \textbf{return} $reach$
    \end{algorithmic}
\end{algorithm}
\end{minipage}

We now describe the approach for SMT-based backward RA. 
For $k\geq 1$, we use $\mathcal{V}^{(-k)}=\{\texttt{x}^{(-k)}_1,\ldots,\texttt{x}^{(-k)}_n\}$ to represent the set of variables encompassing $k^\text{th}$ backward states obtained from $\mathcal{V}^{(0)}$.
The backward version of \eqref{test1} is
\begin{equation}
 Y^{(0)}\wedge \left(\bigwedge_{k=1}^N \mathtt{Im}(\mathcal{V}^{(-k)},\mathcal{V}^{(1-k)})\right) \wedge  X^{(-N)}.   
\label{back_smt}
\end{equation}
Similarly, the one-step version of \eqref{back_smt} can be encoded as 
\begin{equation}
 Y^{(0)} \wedge \mathtt{Im}^N(\mathcal{V}^{(-1)},\mathcal{V}^{(0)}) \wedge X^{(-1)}.   
\label{back_smt_oneshot}
\end{equation}

Algorithms 7 and 8 summarise the backward approach to solve RA via SMT-solving. Line 10 of Algorithm 7 and line 11 of Algorithm 8 are equivalent to the emptiness checking in line 7 of Algorithms 3-4.
\vspace*{-1ex}\\
\begin{minipage}{0.23\textwidth}
\begin{algorithm}[H]
	\footnotesize
    \centering
    \hspace*{-12ex}\textbf{Inputs:} $A\in \Rm{n}{n}$,\\
    \hspace*{-0.5ex}  initial set $X$,\\
    \hspace*{-1ex} target set $Y$,\\
    \hspace*{-8ex} $N\in \mathbb{N}$,\\
    \hspace*{-13ex}\textbf{Output:} boolean
    \caption{\scriptsize SMT-RA (backward)}\label{alg7}
    \begin{algorithmic}[1]
        \State $reach\gets \texttt{false}$
        \State $\mathcal{V}^{(0)}\gets \mathsf{fresh}\_\mathsf{var}(0,n)$
        \State $F\gets \emptyset$ \Comment{{\scriptsize empty vector}}
        \State $F.\mathsf{push}\_\mathsf{back}(Y)$       
        \State $k\gets 1$
        \While{$k\leq N$}
        \State $\mathcal{V}^{(-k)}\gets \mathsf{fresh}\_\mathsf{var}(-k,n)$
        \State $mpl\gets \mathtt{Im}(\mathcal{V}^{(-k)},\mathcal{V}^{(1-k)})$
        \State $F.\mathsf{push}\_\mathsf{back}(mpl)$
        \If{$\mathsf{check}(\wedge  F)\!=\!\texttt{false}$}
        \State \textbf{break}
        \EndIf
        \State $X.\mathsf{subs}(\mathcal{V}^{(1-k)},\mathcal{V}^{(-k)})$
        \State $F.\mathsf{push}\_\mathsf{back}(X)$
        \If{$\mathsf{check}(\wedge F)=\texttt{true}$}
        \State $reach\gets \texttt{true}$
        \State \textbf{break}
        \EndIf
        \State $F.\mathsf{pop}\_\mathsf{back}()$
        \State $k\gets k+1$
        \EndWhile
        \State \textbf{return} $reach$
    \end{algorithmic}
\end{algorithm}
\end{minipage}
\begin{minipage}{0.255\textwidth}
\begin{algorithm}[H]
	\footnotesize
    \centering
    \hspace*{-12ex}\textbf{Inputs:} $A\in \Rm{n}{n}$,\\
    \hspace*{-0.5ex}  initial set $X$,\\
    \hspace*{-1ex} target set $Y$,\\
    \hspace*{-8ex} $N\in \mathbb{N}$,\\
    \hspace*{-13ex}\textbf{Output:} boolean
    \caption{\scriptsize SMT-RA (one-shot backward)}\label{alg8}
    \begin{algorithmic}[1]
        \State $reach\gets \texttt{false}$
        \State $\mathcal{V}^{(0)}\gets \mathsf{fresh}\_\mathsf{var}(0,n)$
        \State $\mathcal{V}^{(-1)}\gets \mathsf{fresh}\_\mathsf{var}(-1,n)$
        \State $F\gets \emptyset$ \Comment{{\scriptsize empty vector}}
        \State $F.\mathsf{push}\_\mathsf{back}(Y)$
        \State $F.\mathsf{push}\_\mathsf{back}(\texttt{true})$ 
        \State $X.\mathsf{subs}(\mathcal{V}^{(0)},\mathcal{V}^{(-1)})$ 
        \State $k\gets 1$
        \While{$k\leq N$}
        \State $F[1]\gets \mathtt{Im}^{k}(\mathcal{V}^{(-1)},\mathcal{V}^{(0)})$
        \If{$\mathsf{check}(\wedge F)=\texttt{false}$}
        \State \textbf{break}
        \EndIf
        \State $F.\mathsf{push}\_\mathsf{back}(X)$
        \If{$\mathsf{check}(\wedge F)=\texttt{true}$}
        \State $reach\gets \texttt{true}$
        \State \textbf{break}
        \EndIf
        \State $F.\mathsf{pop}\_\mathsf{back}()$
        \State $k\gets k+1$
        \EndWhile
        \State \textbf{return} $reach$
    \end{algorithmic}
\end{algorithm}
\end{minipage} 

As we mentioned before, the SMT-based RA of MPL systems is done symbolically in a sense that the SMT formula \eqref{test1} (resp. \eqref{back_smt}) consists of variables from $\mathcal{V}^{(0)}\cup\ldots\cup\mathcal{V}^{(N)}$ (resp. $\mathcal{V}^{(0)}\cup\ldots\cup\mathcal{V}^{(-N)}$). Therefore, if the dimension of matrix in \eqref{mpl} is $n$ then there are $(N+1)\times n$ variables. The number of variables is reduced to $2n$ for the one-shot versions in \eqref{test2} and \eqref{back_smt_oneshot}.  

The performance of the symbolic Algorithms 5-8 depends on the number of ``constraints'' (inequalities and equalities) in \eqref{mpl_smt}. 
If the matrix in $A$ in \eqref{mpl} has $m$ finite elements in each row, then there are $2mn$ constraints.

\section{Computational Benchmarks}
We compare the performance of the SMT-based RA of MPL systems presented in this paper with the existing approach in \citep{DiekyBackward,DiekyForward,Dieky2}. The experiments for both procedures are implemented in C++. For the SMT solver, we use Z3 \citep{Z3}. The computational benchmark has been implemented on an Intel\textregistered{} Xeon\textregistered{} CPU E5-1660 v3, 16 cores, 3.0GHz each, and 16GB of RAM. 

We work with pairs $(n,m)$ where $m\leq n$. For each dimension $n$ (i.e., number of continuous variables), we generate 20 irreducible matrices $A\in \Rm{n}{n}$ with $m$ finite elements in each row, where their values are taken to be between 1 and 20. The locations of the finite elements are chosen randomly. The initial and target sets for each experiment are $X=\{\textbf{x}\in \mathbb{R}^n\mid x_1\geq \ldots\geq x_5\}$ and $Y=\{\textbf{x}\in \mathbb{R}^n\mid x_1\leq \ldots\leq x_5\}$, respectively.  

Table \ref{tab1} (columns 2-5) shows the average running time of the reachability analysis via Algorithms 1 and 3 and of symbolic reachability analysis (SMT-RA) via Algorithms 5 and 7. The $6^\text{th}$ column reports the number of experiments (out of 20) with a $\texttt{true}$ outcome, whilst the last one represents average and maximum completeness threshold, as obtained from the 20 experiments.

\begin{table}[!ht]
\centering
\renewcommand{\tabcolsep}{1mm}
\caption[]{\centering \begin{minipage}[t]{0.7\linewidth}Computational benchmark for sequential reachability analysis of MPL systems\end{minipage}}
\label{tab1}
\begin{scriptsize}
\begin{tabular}{|c|r|r|r|r|c|c|}
\hline
  \multicolumn{1}{|c|}{\multirow{2}{*}{$(n,m)$}}& \multicolumn{2}{c|}{RA}&\multicolumn{2}{c|}{SMT-RA}&\multicolumn{1}{c|}{\multirow{2}{*}{$\#\texttt{true}$}}&\multicolumn{1}{c|}{\multirow{2}{*}{$N^\ast$}}\\\cline{2-5}
       &\multicolumn{1}{c|}{Alg. 1}&\multicolumn{1}{c|}{Alg. 3}  &\multicolumn{1}{c|}{Alg. 5}&\multicolumn{1}{c|}{Alg. 7} &&\\
    \hline 
    $(5,3)$&0.03$s$&0.03$s$&0.02$s$&0.01$s$&7&$\{12.25,30\}$\\\hline
    $(6,3)$&0.31$s$&0.05$s$&0.08$s$&0.01$s$&4&$\{11.20,39\}$\\\hline
    $(7,3)$&5.26$s$&0.47$s$&0.09$s$&0.01$s$&7&$\{10.45,30\}$\\\hline
    $(8,3)$ & $23.89s$ & $3.94s$ & $0.09s$ & $0.01s$ & 10& $\{14.65,49 \}$\\\hline
    $(8,4)$ & $42.14s$ & $11.02s$ & $0.16s$ & $0.01s$ &10 & $\{12.85,21 \}$\\\hline
    $(8,5)$ & $57.84s$ & $21.71s$ & $0.09s$ & $0.01s$ &11 & $\{11.50,33 \}$\\\hline
    $(8,6)$ & $46.42s$ & $40.39s$ & $0.18s$ & $0.01s$ &15 & $\{11.95,20 \}$\\\hline
    $(8,7)$ & $51.28s$ & $28.34s$ & $0.08s$ & $0.01s$ &10 & $\{10.55,22 \}$\\\hline
    $(8,8)$ & $68.51s$ & $40.50s$ & $0.15s$ & $0.02s$ & 13& $\{9.65,30 \}$\\\hline
    $(9,9)$ & $2650.51s$ & $701.29s$ & $0.88s$ & $0.01s$ &9 & $\{ 13.00,25\}$\\\hline
\end{tabular}
\end{scriptsize}
\end{table}

As one can see in Table \ref{tab1}, the SMT-based algorithms are significantly faster than those that explicitly compute reach sets. With regards to the comparison between the forward and backward approaches (for both RA and SMT-RA), the latter seems to be faster. This is likely due to the break condition in line 7 of Algorithm 3 and line 10 of Algorithm 7, which cause the backward algorithms to terminate earlier than the specified step bound $N$ whenever the RA problem has empty solution.  
It should also be noted that the completeness threshold also affects the overall running time. 

Table \ref{tab2} reports the average running times obtained using one-shot approaches over the same tests of Table \ref{tab1} (the last two columns are indeed equal). 
The one-shot strategy improves the running time over its sequential counterpart, particularly in the case of the forward sequential RA algorithms. 
Within the one-shot procedures, again the SMT-based algorithms outperform those sequentially computing the reach sets. 

\begin{table}[!ht]
\centering
\renewcommand{\tabcolsep}{1mm}
\caption[]{\centering \begin{minipage}[t]{0.7\linewidth}Computational benchmark for one-shot reachability analysis of MPL systems\end{minipage}}
\label{tab2}
\begin{scriptsize}
\begin{tabular}{|c|r|r|r|r|c|c|}
\hline
  \multicolumn{1}{|c|}{\multirow{2}{*}{$(n,m)$}}& \multicolumn{2}{c|}{RA}&\multicolumn{2}{c|}{SMT-RA}&\multicolumn{1}{c|}{\multirow{2}{*}{$\#\texttt{true}$}}&\multicolumn{1}{c|}{\multirow{2}{*}{$N^\ast$}}\\\cline{2-5}
       &\multicolumn{1}{c|}{Alg. 2}&\multicolumn{1}{c|}{Alg. 4}  &\multicolumn{1}{c|}{Alg. 6}&\multicolumn{1}{c|}{Alg. 8} &&\\
    \hline 
 $(5,3)$&0.03$s$&0.02$s$&0.01$s$&0.01$s$&7&$\{12.25,30\}$\\\hline
    $(6,3)$&0.22$s$&0.19$s$&0.02$s$&0.01$s$&4&$\{11.20,39\}$\\\hline
    $(7,3)$&1.36$s$&0.91$s$&0.02$s$&0.01$s$&7&$\{10.45,30\}$\\\hline
$(8,3)$ & $9.06s$ & $6.56s$ & $0.02s$ & $0.01s$ & 10& $\{14.65,49 \}$\\\hline
    $(8,4)$ & $13.32s$ & $9.02s$ & $0.02s$ & $0.01s$ &10 & $\{12.85,21 \}$\\\hline
    $(8,5)$ & $20.58s$ & $14.62s$ & $0.02s$ & $0.01s$ &11 & $\{11.5,33 \}$\\\hline
    $(8,6)$ & $27.69s$ & $24.64s$ & $0.02s$ & $0.01s$ &15 & $\{11.95,20 \}$\\\hline
    $(8,7)$ & $32.55s$ & $29.40s$ & $0.02s$ & $0.01s$ &10 & $\{10.55,22 \}$\\\hline
    $(8,8)$ & $42.60s$ & $37.69s$ & $0.02s$ & $0.01s$ & 13& $\{9.65,30 \}$\\\hline
    $(9,9)$ & $843.13s$ & $693.99s$ & $0.03s$ & $0.01s$ &9 & $\{ 13,25\}$\\\hline
\end{tabular}
\end{scriptsize}
\end{table}

The impressive (and almost constant) outcomes of the SMT-RA (Algorithms 6,8) in Table \ref{tab2} suggest to push their scalability to the limit. 
Hence, we provide a computational benchmark for high-dimensional MPL systems in Table \ref{tab3}. 
We focus the benchmark exclusively on one-shot algorithms, as we have seen that sequential algorithms are slower. 
To evenly balance success and failures of RA, we re-define the (dimension of) initial and target sets to be function of the model dimension, as follows: $X=\{\textbf{x}\in \mathbb{R}^n\mid x_1\geq \ldots\geq x_{p}\},Y=\{\textbf{x}\in \mathbb{R}^n\mid x_1\leq \ldots\leq x_{p}\}$ where $p=\frac{n}{2}$.

\begin{table}[!ht]
\centering
\renewcommand{\tabcolsep}{1mm}
\caption[]{\centering \begin{minipage}[t]{0.8\linewidth}Computational benchmark for SMT-based reachability analysis of high-dimensional MPL systems
\end{minipage}}
\label{tab3}
\begin{scriptsize}
\begin{tabular}{|c|r|r|c|c|}
\hline
  \multicolumn{1}{|c|}{\multirow{2}{*}{$(n,m)$}}&\multicolumn{2}{c|}{SMT-RA}&\multicolumn{1}{c|}{\multirow{2}{*}{$\#\texttt{true}$}}&\multicolumn{1}{c|}{\multirow{2}{*}{$N^\ast$}}\\\cline{2-3}
       &\multicolumn{1}{c|}{Alg. 6}&\multicolumn{1}{c|}{Alg. 8}&&\\\hline 
    $(20,10)$&$0.23s$&$0.05s$&8&$\{19.15,44\}$\\\hline
    $(30,15)$&$0.87s$&$0.14s$&5&$\{20.70,30\}$\\\hline
    $(40,20)$&$3.18s$&$0.30s$&2&$\{23.35,47\}$\\\hline
    $(50,25)$&$5.67s$&$0.55s$&1&$\{22.10,50\}$\\\hline
    $(60,30)$&$8.86s$&$1.76s$&3&$\{19.65,34\}$\\\hline
    $(70,35)$&$16.59s$&$3.25s$&1&$\{18.35,37\}$\\\hline
    $(80,40)$&$29.20s$&$6.62s$&0&$\{16.15,25\}$\\\hline
    $(90,45)$&$31.93s$&$12.29s$&2&$\{13.65,21\}$\\\hline
    $(100,50)$&$46.01s$&$21.34s$&5&$\{12.05,14\}$\\\hline
    $(110,55)$&$70.15s$&$43.57s$&4&$\{11.10,12\}$\\\hline
    $(120,60)$&$102.40s$&$70.99s$&2&$\{11.13,13\}$\\\hline
    $(140,70)$&$154.72s$&$92.28s$&4&$\{9.6,11\}$\\\hline
  $(160,80)$&$220.23s$&$222.71s$&6&$\{8.3,10\}$\\\hline
   $(180,90)$&$380.96s$&$539.16s$&11&$\{8,9\}$\\\hline
   $(200,100)$&$682.10s$&$1592.28s$&12&$\{7.35,12\}$\\\hline
\end{tabular}
\end{scriptsize}
\end{table}

Similar to the results in Table \ref{tab2}, Table \ref{tab3} shows that the performance of Algorithm 8 (backward RA) is better than that of Algorithm 6 (forward RA) up to dimension of 140. 
Instead, for larger dimensions the forward RA algorithm outperforms the backward one. 
There are two possible reasons for this outcome. 
First, the larger proportion of $\texttt{true}$ experiments:  
we argue that if the RA problem yields $\texttt{true}$, 
then Algorithm 6 (which performs SMT-checking once) is likely faster than Algorithm 8 (which uses SMT-checking twice for each iteration). 
Second, the smaller values of completeness thresholds also contribute to the relative speedup of the forward algorithm.

Recall that we expect one-shot algorithms to be faster: as an example, for $(n,m)=(50,25)$ the average running time for Algorithm 5 and 7 would be $706.73$ second and $1.33$ second, respectively. Indeed, for the SMT-RA procedures, the one-shot algorithms handle less complex difference logic formulae than the sequential ones: notice that in line 9 of Algorithms 5 and 7, at any iteration $k$ a new formula encompassing the $k^\text{th}$ image of the MPL system is added and sent to the SMT solver; instead, in Algorithms 6 and 8, at every iteration the SMT formula is replaced by a new one, and this is likely to result in simpler formulae. 

\section{Conclusions and Future Work}
This paper has introduced a symbolic approach to solve reachability problems over MPL systems. 
We encode the problems as a formulae in difference logic 
and verify their satisfaction using an SMT solver.

The procedure has been tested on computational benchmarks, 
which have shown a significant improvement over alternative, state-of-the-art techniques. 
Furthermore, the procedure is scalable as it allows to perform reachability analysis of high-dimensional MPL systems. 

As for future research, we are interested to extend the symbolic reachability analysis procedure to \emph{uncertain} MPL systems. 
\begin{ack}
The authors are grateful to Alessandro Cimatti, Andrea Micheli and Mirco Giacobbe for useful comments on SMT. 
The first author is supported by the Indonesia Endowment Fund for Education (LPDP), the last in part by the Alan Turing Institute, London, UK.  
\end{ack}

\bibliography{References}             
                                                   







\end{document}